\documentstyle[12pt]{article}

\title{
\begin{flushright}
{\normalsize Yaroslavl State University\\
             Preprint YARU-HE-94/02\\
             hep-ph/9406347} \\[3cm]
\end{flushright}
Vector Leptoquarks Could Be Rather Light?}

\author{A.V.~Kuznetsov and N.V.~Mikheev\thanks
{E-mail addresses: phth@cnit.yaroslavl.su, physteo@univ.yars.free.net}\\
{\small\it Division of Theoretical Physics, Department of Physics,}\\
{\small\it Yaroslavl State University, Sovietskaya 14,}\\
{\small\it 150000 Yaroslavl, Russian Federation.}}

\date{21 June 1994}

\begin{document}

\maketitle

\begin{abstract}

Some low-energy manifestations of
a minimal extension of the Standard Model based on the
quark-lepton $SU(4)_V \otimes SU(2)_L \otimes G_R$ symmetry of the
Pati-Salam type are analysed. Given this symmetry
a new type of mixing in the quark-lepton interactions is shown to be
required. An additional arbitrariness of the mixing parameters could
allow to decrease noticeably the lower bound on the
leptoquark mass $M_X$ originated from the $\pi$ and $K$ decays
and the $\mu e$ conversion.
The only mixing independent bound emerging from the
cosmological limit on the $\pi^0 \rightarrow \nu \bar{\nu}$ decay
width is $M_X > 18~TeV$.

\end{abstract}

\newpage

Although the Standard Model predictions are in a good agreement with
an experiment now, see e.g. ref.~\cite{A}, a hope for new physics
beyond the Standard Model undoubtedly exists. If the high symmetry
consecutive restoration with an energy increase is assumed, the questions
are pertinent of the next symmetry being restored after the electroweak
one and the next mass scale arising after the $m_W$ scale. We should
like to draw the attention to the alternative which probably has not
been completely studied. It is the Minimal Quark-Lepton Symmetry Model
based on the $SU(4)_V$ group with the lepton number as the fourth
color~\cite{PSm}

\begin{equation}
\left ( \begin{array}{c} u^1 \\ u^2 \\ u^3 \\ \nu \end{array}
\right )_i \, , \qquad \left (
\begin{array}{c} d^1 \\ d^2 \\ d^3 \\ \ell \end{array} \right )_i \, ,
\qquad (i=1,2,3) , \label{eq:q}
\end{equation}

\noindent where the $i$ index labels the fermion generations. The
following attractive features of this extension of the Standard Model
could be specified:
(i) a natural explanation for the quark fractional hypercharge,
(ii) an absence of the proton decay, and
(iii) an indirect evidence that the broken left-right symmetry of the
$SU(2)_L \otimes SU(2)_R$ type could exist.

In the recent paper~\cite{S}
the possible manifestations of an extra $Z'$ boson arising within the
quark-lepton symmetry model have been considered. There appears to be more
exotic object as the charged and colored gauge $X$ boson named leptoquark
which causes the interconversions of quarks and leptons. The tightest
restriction on the leptoquark mass $M_X$ is known~\cite{PDG} to be obtained
from the analyses of $\pi$ and $K$ meson rare decays. The estimations
of the lower bound on the leptoquark mass were carried out in
refs.~\cite{Sha,Des}, however, the mixing in the interactions of the
leptoquark currents was not taken into account there. It can be shown that
such a mixing inevitably occurs in the theory. Really, the mixing in the
quark interaction with the $W$ bosons being depicted by
the Cabibbo-Kobayashi-Maskawa matrix is sure to exist
in Nature. The quark-lepton symmetry necessarily causes the mixing in the
leptoquark interaction, say, with $d,s,b-$ quarks and the charged leptons
$\ell$. On the other hand, it leads at the loop-level to the ultraviolet
divergent non-diagonal transitions $\ell \rightarrow \ell'$ in vacuum
through the quark-leptoquark pair. Consequently, it is necessary for the
renormalizability of the model to include all kinds of mixing at the
tree-level.

Let us suppose that the Standard Model $SU(3)_c \otimes SU(2)_L \otimes U(1)$
extension is realized on the base of the $SU(4)_V \otimes SU(2)_L
\otimes G_R$ group. We assume that the right $G_R$ symmetry will
restore at the appreciably higher mass scale than the $SU(4)_V$ one,
so the explicit form of the $G_R$ group is not adjusted. It is
interesting to note, however, that, if the $G_R$ group is interpreted
as the $U(1)_R$ group of the right hypercharge $Y_R$~\cite{S}, we have
$Y_R = \pm 1$ for the up and down fermions, both quarks and leptons.
It is tempting to interpret this fact as the evidence for the
right hypercharge to be actually the doubled third component of the right
isospin. Hence the $G_R$ group is possible to be the $SU(2)_R$ one.
Further we restrict
ourselves to the consideration of the semi-simple group $SU(4)_V
\otimes SU(2)_L$. Three fermion generations are combined into the \{4,2\}
representations of the type

\begin{equation}
\left ( \begin{array}{c} u^c \\ \nu \end{array} \,
\begin{array}{c} d^c \\ \ell \end{array}
\right )_i , \qquad (i=1,2,3) , \label{eq:d}
\end{equation}

\medskip

\noindent where $c$ is the color index to be further omitted. In the
general case, none of the $\, u,d,\nu,\ell \,$ components is the mass
eigenstate. Due to the identity of three representations~(\ref{eq:d})
they always could be regrouped so that one of the components was
diagonalized with respect to mass. If we diagonalize the charged
lepton mass matrix, the representations~(\ref{eq:d}) can be rewritten
to the form

\begin{equation}
\left ( \begin{array}{c} u \\ \nu \end{array} \,
\begin{array}{c} d \\ \ell \end{array}
\right )_\ell \; = \;
\left ( \begin{array}{c} u_e \\ \nu_e \end{array} \,
\begin{array}{c} d_e \\ e \end{array}
\right ) , \;
\left ( \begin{array}{c} u_\mu \\ \nu_\mu \end{array} \;
\begin{array}{c} d_\mu \\ \mu \end{array}
\right ) , \;
\left ( \begin{array}{c} u_\tau \\ \nu_\tau \end{array} \;
\begin{array}{c} d_\tau \\ \tau
\end{array} \right ) , \label{eq:d2}
\end{equation}

\medskip

\noindent where the indices \, $\ell = e, \mu, \tau$ \, correspond to
the states which are not the mass eigenstates and are included
into the same representations as the charged leptons $\, \ell$

\begin{equation}
\nu_\ell \, = \, {\cal K}_{\ell i} \nu_i , \quad u_{\ell} \, = \,
{\cal U}_{\ell p} u_p , \quad d_{\ell} \, = \, {\cal D}_{\ell n}
d_n . \label{eq:nu1}
\end{equation}

\medskip

\noindent Here \, $\nu_i, u_p, $ and $d_n$ \, are the mass eigenstates

\begin{eqnarray}
\nu_i \, = \, (\nu_1, \, \nu_2, \, \nu_3), \quad u_p \, = \,
(u_1, \, u_2, \, u_3), \, = \, (u, \, c, \, t), \nonumber \\
d_n \, = \, (d_1, \, d_2, \, d_3), \, = \, (d, \, s, \, b),
\label{eq:nu2}
\end{eqnarray}

\medskip

\noindent and ${\cal K}_{\ell i} , \, {\cal U}_{\ell p}$, and
${\cal D}_{\ell n}$ \, are the unitary mixing matrices.

The well-known Lagrangian of the interaction of the charged weak currents
with the $W$ bosons in our notations has the form

\begin{eqnarray}
{\cal L}_W & = & \frac{g}{2 \sqrt 2} \big [
\big ( \bar \nu_{\ell} O_{\alpha} \ell \big ) +
\big ( \bar u_{\ell} O_{\alpha} d_{\ell} \big ) \big ] W^*_{\alpha} +
h.c. \, = \nonumber \\
& = & \frac{g}{2 \sqrt 2} \big [ {\cal K}^*_{\ell i}
\big ( \bar \nu_i O_{\alpha} \ell \big ) + {\cal U}^*_{\ell p} \;
{\cal D}_{\ell n} \;
\big ( \bar u_p O_{\alpha} d_n \big ) \big ] W^*_{\alpha} + h.c.,
\label{eq:Lw}
\end{eqnarray}

\medskip

\noindent where $g$ is the $SU(2)_L$ group constant, and \,$O_{\alpha} \,
= \, {\gamma}_{\alpha}(1-{\gamma}_5)$. The standard Cabibbo-Kobayashi-
Maskawa matrix is thus seen to be $V \, = \, {\cal U}^+ \cal D$.
This is as far as we know about $\cal U$ and $\cal D$ matrices.
The $\cal K$ matrix descriptive of the mixing in the lepton sector has been
the object of intensive experimental investigations in recent years.

Subsequent to the spontaneous $SU(4)_V$ symmetry breaking up to $SU(3)_c$
on the $M_X$ scale six massive vector bosons are separated from the 15-plet
of the gauge fields to generate three charged and colored leptoquarks.
Their interaction with the fermions~(\ref{eq:nu2}) has the form

\begin{equation}
{\cal L}_X \, = \, \frac{g_S(M_X)}{\sqrt 2} \big [
{\cal D}_{\ell n}
\big ( \bar \ell \gamma_{\alpha} d^c_n \big ) +
\big ( {\cal K^+ \; \cal U} \big )_{i p}
\big ( \bar{\nu_i} \gamma_{\alpha} u^c_p \big ) \big ] X^c_{\alpha} +
h.c. \,,
\label{eq:Lx}
\end{equation}

\noindent where the color index $'c'$ is written once again. The constant
\, $g_S(M_X)$ \, can be expressed in terms of the strong coupling constant
\, $\alpha_S$ \, at the leptoquark mass scale
$M_X, \quad g_S^2(M_X)/4 \pi = \alpha_S(M_X)$.

If the momentum transferred is \, $q \ll M_X$, \, then the Lagrangian
{}~(\ref{eq:Lx}) in the second order leads to the effective four-fermion
vector-vector interaction of quarks and leptons. By using the Fiertz
transformation, lepton-current-to-quark-current terms of the scalar,
pseudoscalar, vector and axial-vector types may be separated in the
effective Lagrangian. Let us note that the construction of the effective
lepton-quark interaction Lagrangian requires taking account of the QCD
corrections estimated by the known technique~\cite{Vai,Vys}.
In our case the leading log approximation $ln(M_X/\mu) \gg 1$ with
$\mu \sim 1~GeV$ to be the typical hadronic scale is quite applicable.
Then the QCD correction amounts to the appearance of the magnifying factor
$Q(\mu)$ at the scalar and pseudoscalar terms

\begin{equation}
Q(\mu) \, = \, \left ( \frac{\alpha_S(\mu)}
{\alpha_S(M_X)} \right )^{4/\bar b} \, .
\label{eq:Qmu}
\end{equation}

\noindent Here $\alpha_S(\mu)$ is the effective strong coupling constant
at the hadron mass $ \, \mu \,$ scale,
$\; \bar b \, = \, 11 \, - \, \frac {2}{3} \bar n_f, \; \bar n_f $
is the averaged number of the quark
flavors at the scales $\mu^2 \le q^2 \le M_X^2$. If the condition
$M_X^2 \gg m_t^2$ takes place, then we have $\, \bar n_f \, \simeq \, 6$,
and $\bar b \, \simeq \, 7$.

Let us investigate the contribution of the leptoquark interaction~
(\ref{eq:Lx}) to the low-energy processes to establish the bounds on the
model parameters from existing experimental limits.
As the analysis shows, the tightest restrictions on the leptoquark mass
$M_X$ and the mixing matrix $\cal D$ elements
can be obtained from the experimental data on rare $\pi$ and $K$ decays and
$\mu^- \rightarrow e^-$ conversion in nuclei. They are represented in the
left part of table 1.

In the description of the interactions of $\pi$ and $K$ mesons it is
sufficient to take the scalar and pseudoscalar terms only.
As we shall see later, these terms acquire in addition to the QCD corrections
an extra enhancement at the amplitude by the small quark current masses.
The corresponding part of the effective Lagrangian takes the form

\begin{eqnarray}
\Delta \, {\cal L}_{eff} & = & - \frac{2 \pi \alpha_S(M_X)}{M_X^2} \,
Q(\mu) \; \big [ {\cal D}_{\ell n} \big ( {\cal U^+ \cal K} \big )_{pi}
\big ( \bar{\ell} \gamma_5 \nu_i \big ) \big ( \bar u_p \gamma_5 d_n
\big ) + h.c. + \nonumber \\
\hspace{3mm}
& + & {\cal D}_{\ell n}
{\cal D}^*_{\ell' n'} \big ( \bar \ell \gamma_5 \ell' \big )
\big ( \bar d_{n'} \gamma_5 d_n \big ) + \nonumber \\
\hspace{5mm}
& + & \big ( {\cal K^+ \; \cal U} \big )_{i p}
\big ( {\cal U^+ \; \cal K} \big )_{p'i'}
\big ( \bar{\nu_i} \gamma_5 \nu_{i'} \big )
\big ( \bar u_{p'} \gamma_5 u_p \big ) \; - \; (\gamma_5 \rightarrow 1)
\big ] . \label{eq:Lef}
\end{eqnarray}

\medskip

\noindent These interactions will contribute to the rare $\pi$
and $K$ meson decays strongly suppressed
in the Standard Model.

One can easily see that the leptoquark contribution to the $\pi
\rightarrow e \nu$ decay is not suppressed by the electron mass
in contrast to the $W-$contribution. The corresponding part of the
amplitude could be represented in the form

\begin{equation}
{\Delta \cal M}^X_{\pi e \nu} \, = \, - \, \frac{2 \pi \alpha_S(M_X)}{M^2_X} \,
{\cal D}_{e d} \, {\cal U}^*_{\ell u} \;
\frac{f_{\pi} \; m^2_{\pi} \; Q(\mu)}{m_u(\mu) + m_d(\mu)} \,
\big ( \bar e \gamma_5 \nu_{\ell} \big ) ,
\label{eq:MpiX}
\end{equation}

\noindent where $f_{\pi} \simeq 132~MeV$ is the $\, \pi \ell \nu \,$ decay
constant, $\, m_{u,d}(\mu) \,$ are the running quark masses
at the \, $\mu$ \, scale. Let us note that the ratio $Q(\mu)/m(\mu)$
is the renormalization group invariant, since the $Q(\mu)$
function~(\ref{eq:Qmu}) determines also the law of the quark mass running.
To the $\mu \simeq 1~GeV$ scale there correspond the well-known quark
current masses $m_u \simeq 4~MeV, m_d \simeq 7~MeV$ and
$m_s \simeq 150~MeV$, see e.g. refs.~\cite{Gas,W}.
Taking into account the interference of the amplitude~
(\ref{eq:MpiX}) and the known $W-$
exchange amplitude we get the following expression for
the decay width ratio $\Gamma (\pi \rightarrow e \nu)
/ \Gamma (\pi \rightarrow \mu \nu) \equiv R $

\begin{equation}
R \; = \; R_W \big [ 1 \; - \;
\frac{2 \sqrt 2 \pi \; \alpha_S(M_X) \; m^2_{\pi} \; Q}
{G_F M^2_X m_e (m_u + m_d)} \; Re \big ( \frac{{\cal D}_{e d}
{\cal U}^*_{e u}}{V_{u d}} \big ) \big ],
\label{eq:R}
\end{equation}

\noindent where $R_W = (1.2352 \pm 0.0005)~\cdot~10^{-4}$ is the value of the
ratio in the Standard Model~\cite{Mar}. Using the combined results on $R$ of
two recent experiments at PSI and TRIUMF~\cite{Bri}
we get the following lower bound on the leptoquark
mass at the 90 $\%$ C.L., see line 1 of table 1.
In the paper~\cite{Sha} an attempt was also made to estimate
the leptoquark mass from the $R$ ratio, however, an additional
assumption was taken there on the $SU(4)$ group constant to be equal
to the weak constant $g$, and the QCD corrections were not considered.

An amplitude of the process $K^0_L \rightarrow e^- \mu^+$ can be found
similarly to eq.~(\ref{eq:MpiX}) to be

\begin{equation}
{\cal M}^X_{K e \mu} \, = \, - \, \frac{\sqrt 2 \pi \alpha_S(M_X) \;
f_K \; m^2_K \; Q}{M^2_X \, (m_s + m_d)} \;
\big ( {\cal D}_{e d}
{\cal D}^*_{\mu s} \; + \;
{\cal D}_{e s}
{\cal D}^*_{\mu d} \big ) \;
\big ( \bar e \gamma_5 \mu \big ) .
\label{eq:MKX}
\end{equation}

\noindent where $f_K \simeq 160~MeV$ is the $\, K \ell \nu \,$ decay
constant.
In place of the estimation of the leptoquark mass bound
$M_X > 350~TeV$ obtained in ref.~\cite{Des}, where the mixing in the
leptoquark interaction was not taken into account, we find
from the precised experimental data~\cite{Ari} the bound,
see line 5 of table 1.

Within recent years the experimental limit on the
$Br(K^0_L \rightarrow \mu^+ \mu^-)$ value was noticeably
lowered to became close to the unitary limit $Br_{abs} =
6.8 \cdot 10^{-9}$.
Thus the effective leptoquark contribution to $Br(K^0_L \rightarrow \mu^+
\mu^-)$ isn't likely to be larger than $(1 \div 2) \cdot 10^{-9}$.
The process amplitude can be easily obtained from eq.~(\ref{eq:MKX})
by replacing $e \rightarrow \mu$.

An amplitude of one more rare $K^0_L$ decay into electron and positron
through the intermediate leptoquark could be also obtained from
eq. (\ref{eq:MKX}) by replacing $\mu \rightarrow e$. The limits on the
model parameters in these cases are represented in lines 4,6 of table 1.

Among the charged $K^+$ meson rare decays which go at the tree level
in this model, the tightest limits are established on the decays
$K^+ \rightarrow \pi^+ \mu^- e^+$~\cite{Dia} and
$K^+ \rightarrow \pi^+ \mu^+ e^-$~\cite{Lee}. An amplitude of the process
$K^+ \rightarrow \pi^+ \mu^+ e^-$ can be written in the form

\begin{equation}
{\cal M}^X_{K \pi \mu e} \, = \,- \frac{2 \pi \alpha_S(M_X)}{M^2_X} \;
\frac{f^0_+(q^2) \, (m^2_K - m^2_{\pi}) + f^0_-(q^2) \, q^2}{m_s - m_d}
\; Q\; {\cal D}_{e d} {\cal D}^*_{\mu s} \; \big ( \bar e \mu \big ) .
\label{eq:MKpi}
\end{equation}

\noindent Here $q$ is the four-momentum of the lepton pair and $f^0_{+,-}$
are the known form factors of the $K^0_{\ell 3}$ decay. An amplitude of
the $K^+ \rightarrow \pi^+ \mu^- e^+$ decay can be obtained from
eq.~(\ref{eq:MKpi}) by interchanging $\, e \,$ and $\mu$. The bounds
on the model parameters arising from these decays contain the same
matrix elements as from the $K^0_L \rightarrow e \mu$ decay but individually,
see lines 2,3 of table 1.

A low-energy process under an intensive experimental searches, where the
leptoquark could manifest itself is the $\mu e$ conversion in nuclei.
The coherent $\mu e$ conversion is the most convenient for the observation
when the nucleus remains in the ground state, and consequently the electrons
are monoenergetic, with the maximum possible energy $\simeq m_{\mu}$.
An effective Lagrangian for the coherent $\mu e$ conversion contains the
scalar and vector quark currents only. In the model under discussion it
has the form

\begin{equation}
\Delta \, {\cal L}_{\mu e} \,= \, - \frac{2 \pi \alpha_S(M_X)}{M_X^2} \,
{\cal D}_{e d} {\cal D}^*_{\mu d} \big [{1 \over 2} \big ( \bar e
\gamma_{\alpha} \mu \big )
\big ( \bar d \gamma_{\alpha} d \big ) \, - \,
\big ( \bar e \mu \big )
\big ( \bar d d \big ) Q(\mu) \big ] .
\label{eq:Lmue}
\end{equation}

\noindent Using the calculation technique of ref.~\cite{Sha79} we estimate
the branching ratio of the $\mu e$ conversion in titanium
for the interaction of the type~(\ref{eq:Lmue}).
The result allows us to establish the bound on the model parameters,
see line 7 of table 1,
on the base of the experimental data~\cite{Ah}.

One can see from table 1 that the restrictions on the model parameters
contain the elements of the
unknown unitary mixing matrices $\cal D$ and $\cal U$, which are connected
by the condition ${\cal U}^+ {\cal D} = V$ only.
Thus the possibility is not excluded, in principle, when the bounds obtained
did not restrict $\, M_X \,$ at all, e.g. if the elements ${\cal D}_{e d}$
and ${\cal D}_{\mu d}$ were rather small. It would correspond to the
connection of the $\tau$ lepton largely with $d$ quark in the ${\cal D}$
matrix, and electron and muon with $s$ and $b$ quarks.
In general, it is not contradictory to anything even if appears unusual.
In this case a leptoquark could give more noticeable
contribution to the flavor-changing decays of $\tau$ lepton and
$\eta, \eta', \Phi$ and
$B$ mesons. However, a relatively poor accuracy of these data doesn't
yet allow to restrict the parameters essentially.

We could find only one occasion when the mixing-independent lower bound
on the leptoquark mass arises, namely, from the decay
$\pi^0 \rightarrow \nu \bar \nu$. In the paper~\cite{Lam} the cosmological
estimation of the width of this decay was found

\medskip

$Br(\pi^0 \rightarrow \nu \bar \nu ) < 2.9 \cdot 10^{-13}$ .

\medskip

\noindent Within the Standard Model this value is proportional to $m^2_\nu$.
The process is also possible through the leptoquark mediation, without the
suppression by the smallness of neutrino mass. The process amplitude has
the form

\begin{equation}
{\cal M}^X_{\pi \nu \nu} \, = \, \frac{\pi \alpha_S(M_X) \; f_{\pi} \;
m^2_{\pi} \; Q}{\sqrt 2 \; M^2_X \, m_u} \;
({\cal K}^+{\cal U})_{i u} \,
({\cal U}^+{\cal K})_{u j} \;
\big ( \bar \nu_i \gamma_5 \nu_j \big ) ,
\label{eq:Mnu}
\end{equation}

\noindent On summation over all neutrino species $i,j$ the decay probability
is mixing-independent. As a result the bound on the leptoquark mass occurs

\begin{equation}
M_X > 18~TeV.
\label{eq:X4}
\end{equation}

\bigskip

In conclusion, we have analysed in detail the experimental data
on rare $\pi$ and $K$ decays and $\mu e$ conversion and we
have found the restrictions on the vector leptoquark mass to contain the
elements of an unknown mixing matrix $\cal D$. The only mixing independent
bound (\ref{eq:X4}) arises from the cosmological estimations.

In our opinion, possible experimental manifestations of the considered
minimal quark-lepton symmetry model would be an object of further
methodical studies. For example, the search of possible leptoquark evidence
in the $p \bar p$ collider high-energy experiments via the reactions
$d \bar d \rightarrow e^+ \mu^-, \; e^- \mu^+$ could be of interest.
On the other hand,
further searches of flavor-changing decays of $\tau$ lepton and
$\eta, \eta', \Phi$ and $B$ mesons are desirable.

\bigskip

{\bf Acknowledgments}

\bigskip

The authors are grateful to
L.B.~Okun, V.A.~Rubakov, K.A.~Ter-Martirosian and A.D.~Smir\-nov
for fruitful discussions.

\newpage

\begin{table}[h]
\caption{The bounds on the leptoquark mass and mixing matrix
elements from the experimental limits on the branching ratios of
of various processes.}

\vspace{10mm}

\begin{center}
\begin{tabular}{cccc}\hline
No. & Experimental limit & Ref. & Bound \\ \hline \\
\bigskip
1 & $\frac{\mbox{\normalsize $\Gamma(\pi \rightarrow e \nu)$}}
{\mbox{\normalsize $\Gamma(\pi \rightarrow \mu \nu)$}}
= (1.2310 \pm 0.0037) \cdot 10^{-4}$ & \cite{Bri} &
$\frac{\mbox{\normalsize $M_X$}}
{\mbox{\normalsize $|Re({\cal D}_{e d} {\cal U}_{e u}^*/V_{ud})|^{1/2}$}} \,
> \, 210~TeV$ \\
\bigskip
2 & $Br(K^+ \rightarrow \pi^+ \mu^- e^+ ) < 7 \cdot 10^{-9}$ &
\cite{Dia} &
$\frac{\mbox{\normalsize $M_X$}}
{\mbox{\normalsize $|{\cal D}_{e s} {\cal D}^*_{\mu d} |^{1/2}$}} \, > \,
50~TeV$ \\
\bigskip
3 & $Br(K^+ \rightarrow \pi^+ \mu^+ e^- ) < 2.1 \cdot 10^{-10}$ &
\cite{Lee} &
$\frac{\mbox{\normalsize $M_X$}}
{\mbox{\normalsize $|{\cal D}_{e d} {\cal D}^*_{\mu s} |^{1/2}$}} \, > \,
120~TeV$ \\
\bigskip
4 & $Br(K^0_L \rightarrow \mu^+ \mu^-) = (7.3 \pm 0.4) \cdot 10^{-9}$ &
\cite{PDG} &
$\frac{\mbox{\normalsize $M_X$}}
{\mbox{\normalsize $|Re({\cal D}_{\mu d} {\cal D}_{\mu s}^*)|^{1/2}$}} \,
> \, 500 \div 600~TeV$ \\
\bigskip
5 & $Br(K^0_L \rightarrow \mu e ) < 3.3 \cdot 10^{-11}$ &
\cite{Ari} &
$\frac{\mbox{\normalsize $M_X$}}
{\mbox{\normalsize $|{\cal D}_{e d} {\cal D}^*_{\mu s} \; + \;
{\cal D}_{e s} {\cal D}^*_{\mu d} |^{1/2}$}} \, > \, 1200~TeV$ \\
\bigskip
6 & $Br(K^0_L \rightarrow e^+e^- ) < 5.3 \cdot 10^{-11}$ &
\cite{Ari} &
$\frac{\mbox{\normalsize $M_X$}}
{\mbox{\normalsize $|Re({\cal D}_{e d} {\cal D}^*_{e s})|^{1/2}$}} \, > \,
1400~TeV$ \\
\bigskip
7 & $\frac{\mbox{\normalsize $\Gamma(\mu^-Ti \rightarrow e^-Ti)$}}
{\mbox{\normalsize $\Gamma(\mu^-Ti \rightarrow capture)$}}
< 4.6 \cdot 10^{-12}$ &
\cite{Ah} &
$\frac{\mbox{\normalsize $M_X$}}
{\mbox{\normalsize $|{\cal D}_{e d} {\cal D}^*_{\mu d} |^{1/2}$}} \, > \,
670~TeV$ \\
\hline
\end{tabular}
\end{center}

\vspace{80mm}

\end{table}

\end{document}